\documentclass[useAMS,usenatbib]{mn2e}
\usepackage{amsmath,amssymb,mathrsfs,graphicx,float,latexsym}
\usepackage{epstopdf,ragged2e}
\usepackage{hyperref}

\newcommand{\Po}{P_{\text{orb}}}
\newcommand{\fGW}{f_{\text{GW}}}
\newcommand{\Mc}{M_{\text{comp}}}
\newcommand{\Ms}{M_{\star}}
\newcommand{\Rs}{R_{\star}}
\newcommand{\nuS}{\nu_{\text{spin}}}

\begin{document}
\title[UCXBs as dual-line GW sources]{Ultra-compact X-ray binaries as dual-line gravitational-wave sources}
\author[A. G.~Suvorov]{A. G.~Suvorov\thanks{E-mail: arthur.suvorov@tat.uni-tuebingen.de}\\Theoretical Astrophysics, Eberhard Karls University of T{\"u}bingen, T{\"u}bingen, D-72076, Germany}

\date{Accepted ?. Received ?; in original form ?}

\pagerange{\pageref{firstpage}--\pageref{lastpage}} \pubyear{?}

\maketitle
\label{firstpage}

\begin{abstract}
\noindent{By virtue of their sub-hour orbital periods, ultra-compact X-ray binaries are promising sources for the space-borne gravitational-wave interferometers LISA, Taiji, and TianQin. Some of these systems contain a neutron star primary, whose spin period can be measured directly via pulse timing, or indirectly through rotational modulations of burst phenomena. It is pointed out here that since actively accreting stars, with spin frequencies in the hundreds of Hz, may continuously emit appreciable gravitational waves due to the presence of accretion-built mountains, toroidal magnetic fields, and/or $r$-mode oscillations, such binaries are also candidate sources for ground-based interferometers. Two Galactic systems (4U 1728--34 and 4U 1820--30) are identified as being potentially detectable by both LISA and aLIGO simultaneously: a dual-line detection of this sort could provide percent-level constraints on the mass, radius, and internal magnetic field strength of the neutron star. With the Einstein Telescope, we find that at least four of the known ultra-compact binaries become dual-line visible.}
\end{abstract}


\section{Introduction} 

Ultra-compact X-ray binaries (UCXBs) are tight (orbital periods $\Po \lesssim$ 1 hr) systems involving a compact object and a low-mass $(M_{\text{comp}} \ll M_{\odot})$ companion. While their genealogy is not fully understood, evolutionary models suggest they arise as the (possibly detached) end state of low-mass X-ray binaries (LMXBs), that have gradually compactified through orbital angular-momentum losses due to a combination of magnetic braking, gravitational-wave (GW) emission, and mass transfers \citep{savon86,post14,chen16,seng17}. The companion mass is gradually stripped away over time, leaving behind a neutron star (NS) or black hole in a close binary with a non-degenerate He star or a O/Ne/Mg, C/O, or He white dwarf (WD) \citep{rap82,bild04}. Due to their short orbital periods, UCXBs are strong candidate sources for space-based GW observatories, most notably the Laser Interferometer Space Antenna (LISA), {Taiji, and TianQin} \citep{nele01,tauris18,luo20,huang20,chen21}. In particular, circular binary systems emit GWs at exactly twice the orbital frequency \citep{finn00} which, for UCXBs, lies in the $\sim$ mHz band and coincides with the peak sensitivity of LISA \citep{rob19}. 

Theoretical calculations based on the assumption that the companion fills its Roche lobe \citep{pac71,rap87}, together with evolutionary modelling \citep{hein13,tauris18} {[which incidentally indicates that there may be up to $\approx 320$ UCXB-LISA sources in the Galaxy from the NS--main-sequence channel alone \citep{chen20}]}, suggest that companion masses in UCXBs are in an approximately one-to-one relationship with the orbital periods. {A measurement of GWs from these systems, which would allow for precise determinations of the chirp mass, may then strongly constrain the mass of the NS primary; for NS-WD binaries in particular, the NS mass may be calculable to within a few percent \citep{tauris18} [see also Table 1 in \cite{chen20}].} This prospect becomes especially interesting for sources that exhibit thermonuclear activity, since the apparent emitting area and the peak flux achieved during photospheric bursts can be used to independently constrain the mass-radius relationship of the star \citep{van79,oz09}. For example, by monitoring X-ray bursts from the NS in 4U 1820--30, \cite{guv10} found $M_{\star} = 1.58 \pm 0.06 M_{\odot}$ and $R_{\star} = 9.1 \pm 0.4 \text{ km}$ at the $1 \sigma$ level for the NS mass and radius, respectively. If combined with an independent, percent-level detection of $M_{\star}$, this would place tight constraints on the nuclear equation of state (EOS) governing NS matter.

As noted by \cite{tauris18}, UCXBs become an even more promising avenue for studying NS physics when one considers that rapidly rotating NSs may emit appreciable \emph{continuous GWs}\footnote{{While orbital motion also continuously generates GWs, we use the phrase \emph{continuous GWs} in this paper to refer exclusively to persistent radiation from the NS itself.}}. For NSs spun-up by accretion torques, it remains an observational puzzle as to why their spin frequencies seem to be capped at a maximum value $\nu_{\text{obs}} \lesssim 650 \text{ Hz}$ \citep{chak03,pat10}, well below the Keplerian break-up limit. A proposed solution to this problem is that, in addition to magnetic braking \citep{ghosh79} and the magnetospheric centrifugal-barrier \citep{ill75}, GW back-reaction may act to stall the spin-up \citep{bild98,hmount15,glamp18}. For systems with high X-ray luminosities, the NS may possess a large mass quadrupole moment due to the formation of an accretion-built magnetic mountain \citep{payne04,melp05} or toroidal magnetic field within the stellar interior \citep{cutler02,lander13}, and radiate at twice the spin frequency \citep{watts08}. Large, current-type quadrupoles may also exist in rapidly rotating systems through $r$-mode oscillations \citep{ands99,and01}; the system is expected to radiate at just over four-thirds of the spin frequency in this case \citep{lock03,and14}. UCXBs may thus be sources of continuous GWs in the frequency range $10^{2} \lesssim \fGW/\text{Hz} \lesssim 10^{3}$ where ground-based interferometers, such as the Advanced Laser Interferometer Gravitational-Wave Observatory (aLIGO) or the upcoming Einstein Telescope (ET), are most sensitive. 

In this Article we examine the possibility of detecting GWs from UCXBs in two distinct channels: $\sim$ mHz-range GWs from the orbital motion (detectable by LISA), and $\lesssim$ kHz-range GWs from the NS primary (detectable by aLIGO or ET). Under certain assumptions (discussed in detail in Sections 2 and 3), it is shown that 4U 1728--34 and 4U 1820--30 (both of which are active X-ray bursters) present two promising candidates for `dual-line' detection, though several more may come into view with the advent of ET. 

\section{UCXBs as LISA sources}

Averaging over orbital orientations and polarizations, a general binary system emits GWs at a frequency $\fGW = 2/ \Po$ with intrinsic amplitude \citep{finn00} 
\begin{equation} \label{eq:lisa0}
h^{\text{orb}}_{0} \approx 5.1 \times 10^{-23} \left( \frac{1 \text{ hr}} {\Po} \right)^{2/3} \left( \frac {\mathcal{M}} {1 M_{\odot}} \right)^{5/3} \left( \frac {10 \text{ kpc}} {d} \right),
\end{equation}
where $\mathcal{M}$ and $d$ denote the chirp mass and source distance, respectively. For most UCXB systems, the orbital decay is gradual ($|\dot{P}_{\text{orb}}| \ll 10^{-10}$) and the GW signal is essentially monochromatic. This allows for an accumulation of signal power in a GW detector over many orbital cycles $N_{\text{cycles}}$. In particular, for observation time $T$, a characteristic strain $h_{c}$ can be estimated through $h_{c} \approx \sqrt{N_{\text{cycles}}} \sqrt{2} h^{\text{orb}}_{0} = 2 \sqrt{T/P_{\text{orb}}} h^{\text{orb}}_{0}$ \citep{finn00,tauris18}. A longer observation time not only augments the effective GW strain, but also helps to reduce the `confusion' noise associated with unresolved Galactic sources [see Table 1 of \cite{rob19}]. For an observation time of $T = 4$ yr, which drastically reduces the Galactic confusion noise around $\sim$ mHz GW frequencies (precisely where UCXBs lie), the characteristic strain associated with a binary reads 
\begin{equation} \label{eq:lisahc}
h_{c} \approx 1.9 \times 10^{-20} \left( \frac{1 \text{ hr}} {\Po} \right)^{7/6} \left( \frac {\mathcal{M}} {1 M_{\odot}} \right)^{5/3} \left( \frac {10 \text{ kpc}} {d} \right).
\end{equation}

In expressions \eqref{eq:lisa0} and \eqref{eq:lisahc}, the chirp mass is defined as
\begin{equation} \label{eq:chirp}
\mathcal{M} = \frac {\left( \Ms \Mc \right)^{3/5}} { \left(\Ms + \Mc \right)^{1/5}},
\end{equation}
which, given $M_{\star}$, can be evaluated using the (observed) binary mass function $f_{\text{X}} = (\Mc \sin i)^{3} (M_{\star} + \Mc)^{-2}$ [e.g., \cite{mark02}], for orbital inclination angle $i$. For systems where $\Mc$ is not well-constrained (e.g., when the angle $i$ is totally unknown), the mass of the donor star can instead be estimated from the \cite{rap87} relation [see also \cite{chen20}]
\begin{equation} \label{eq:chenform}
M_{\text{comp}} \approx 0.013 \left( \frac{ 1 \text{ hr}} {P_{\text{orb}}} \right) M_{\odot},
\end{equation}
derived by assuming that the hydrogen-poor companion fills its Roche lobe and is well-described by the EOS pertaining to a degenerate gas of non-relativistic electrons (as for a WD). However, expression \eqref{eq:chenform} should be handled with care since the companion makeup is seldom known \citep{rap82,bild04}. {In general, an accurate determination of the chirp mass requires a high signal-to-noise ratio (SNR). Using expression (26) in \cite{rob19}, the orientation- and polarisation-averaged SNR for a binary system can be estimated as}
\begin{equation}
\text{SNR}(\fGW) \approx \frac {8 \pi^{2/3} G^{5/3} \mathcal{M}^{5/3} \fGW^{2/3}} {\sqrt{5} c^4 d} \sqrt{ \frac {T} {S_{n}(\fGW)}},
\end{equation}
{for detector noise power spectral density $S_{n}$. For a 4-year observation of a NS-WD binary with primary mass $M_{\star} = 1.6 M_{\odot}$ at a distance $d = 1 \text{ kpc}$, we find $\text{SNR}(10^{-3}) \approx 3$ (i.e., $P_{\text{orb}} = 33$ min) while $\text{SNR}(3 \times 10^{-3}) \approx 145$ ($P_{\text{orb}} = 11$ min) for LISA. For SNRs $\gtrsim 50$, \cite{tauris18} estimates that a measurement error on $\mathcal{M}$ of $\lesssim 1\%$ is achievable for various UCXB systems.}

\begin{table*}
\centering
\caption{Observed and derived properties related to the orbital and accretion dynamics of the UCXBs considered here. Companion masses $\Mc$ are quoted either (in order of decreasing robustness) from observational upper-limits (asterisks), or the \protect\cite{rap87} relation \eqref{eq:chenform} (daggers). X-ray luminosities $L_{X}$ are determined via the peak flux during outbursts in the case of sources exhibiting type I X-ray bursts.}
\hspace{-1.2cm}\begin{tabular}{ccccc}
 \hline
 \hline
Source & $P_{\text{orb}}$ (s) & $\Mc$ ($\times 10^{-2} M_{\odot}$) & $L_{X}$ ($ \times10^{36} \text{ erg s}^{-1}$) & $d$ (kpc) \\
\hline
4U 1820--30$^{\text{a}}$ & 685 & $6.89^{\dag}$ & 57 & 7.6 \\
IGR J16597--3704$^{\text{b}}$ & 2758 & $1.71^{\dag}$ & 6.5 & 9.1 \\
4U 1728--34$^{\text{c}}$ & 646 & $7.30^{\dag}$ & 5.0 & 5.1 \\
HETE J1900.1--2455$^{\text{d}}$ & 4995 & $8.5^{*}$ & 3.7 & 4.3 \\
XB 1916--053$^{\text{e}}$ & 3005 & $10.1^{*}$ & 6.6 & 8.4 \\
XTE J1807--294$^{\text{f}}$ & 2404 & $2.2^{*}$ & $\lesssim$ 13 & $\sim$ 8 \\
4U 1915--05$^{\text{g}}$ & 3027 & $\sim 10^{*}$ & $\sim 6$ & 9.3 \\
4U 0614+091$^{\text{h}}$ & 3060 & $1.54^{\dag}$ & 3.4 & 3.2 \\
4U 0513--40$^{\text{i}}$ & 1020 & $4.63^{\dag}$ & 7.4 & 12 \\
4U 1850--087$^{\text{j}}$ & 1236 & $3.82^{\dag}$ & 1.7 & 8.2 \\
XTE J1751--305$^{\text{k}}$ & 2545 & $3.5^{*}$ & $<$ 18.3 & $>$ 7 \\
XTE J0929--314$^{\text{l}}$ & 2615 & $\sim 3^{*}$ & $<$ 13 & $\gtrsim$ 7.4 \\
\hline
\hline
\end{tabular} 
\\
\footnotesize{$\vphantom{\text{a}}^{\text{a}}$ \cite{guv10,rev11,chen20}. $\vphantom{\text{b}}^{\text{b}}$ \cite{sanna18}. $\vphantom{\text{c}}^{\text{c}}$ \cite{gal10,erg11} [though cf. \cite{vin20}]. $\vphantom{\text{d}}^{\text{d}}$ \cite{fal07,ele08}. $\vphantom{\text{e}}^{\text{e}}$ Sometimes called 4U 1916-05; \cite{church97,iar20}. $\vphantom{\text{f}}^{\text{f}}$ \cite{falan05,rigg08}. $\vphantom{\text{g}}^{\text{g}}$ \cite{grind88,zhang14}. $\vphantom{\text{h}}^{\text{h}}$ Sometimes called H 0614+091; \cite{rev11,saz20}. $\vphantom{\text{i}}^{\text{i}}$ \cite{rev11,chen20}. $\vphantom{\text{j}}^{\text{j}}$ \cite{homer96,rev11}. $\vphantom{\text{k}}^{\text{k}}$ \cite{mark02,gier05,rigg11}. $\vphantom{\text{l}}^{\text{l}}$ \cite{gal02,marin17}.}
\label{tab:orbdata}
\end{table*}

While $\lesssim 40$ UCXB candidates have been identified to date \citep{saz20}, many of these systems involve black hole primaries, have orbital periods with large uncertainties, or contain slowly-spinning NSs. We focus on 12 NS-primary systems for which the orbital periods are especially short (so as to be potentially detectable by LISA; see below) and/or are such that the properties of the NS primary are favourable for detection by ground-based interferometers (see Sec. 3). Table \ref{tab:orbdata} lists the orbital periods $P_{\text{orb}}$, the estimated or measured companion masses $\Mc$ (see Table caption), X-ray luminosities $L_{X}$ (either persistent or associated with peak fluxes during outbursts for flaring sources), and distances $d$ for the UCXBs considered in this work. All sources have orbital periods $< 1$ hr, with the exception of HETE J1900.1--2455 ($P_{\text{orb}} = 1.39$ hr); we include this latter source because it has been identified as a candidate for continuous GW detection by \cite{hask15} and \cite{hmount15}. Note that the chirp mass \eqref{eq:chirp} is also sensitive to the NS mass, for which we take $M_{\star} = 1.6 M_{\odot}$ unless direct estimates have been made (see Sec. 3 and Table \ref{tab:nsdata}).

\begin{figure}
\hspace{-0.3cm}\includegraphics[width=0.5\textwidth]{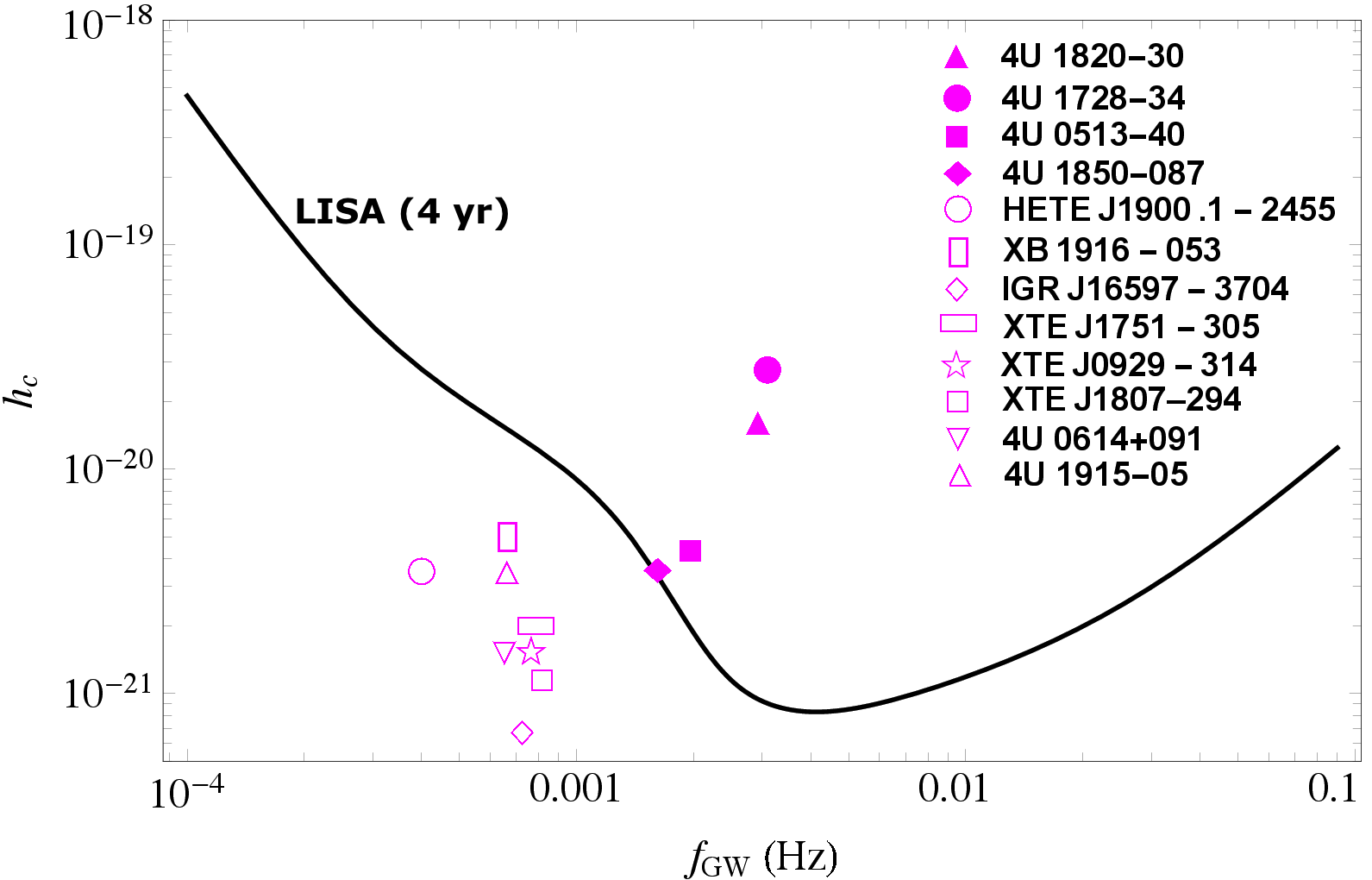}
\caption{Characteristic strain $h_{c}$ as a function of GW frequency associated with the orbital motion for 12 UCXBs (see plot legends), whose system parameters are detailed in Tab. \ref{tab:orbdata}, assuming a 4 year observation period [equation \eqref{eq:lisahc}]. Overplotted is an analytic approximation to the LISA noise curve; see Table 1 of \protect\cite{rob19}.} \label{fig:lisastrains}
\end{figure}

From the data collated in Tab. \ref{tab:orbdata}, we can estimate the characteristic (orbital) GW strain \eqref{eq:lisahc}, relevant for a 4 year observation run with LISA. Figure \ref{fig:lisastrains} plots $h_{c}$ for each source together with the LISA noise curve from \cite{rob19} as functions of the GW frequency $\fGW = 2/ \Po$. We see that four of the sources we consider {[namely 4U 1820--30 (SNR $\approx$ 17), 4U 1728--34 (SNR $\approx$ 30), 4U 0513--40 (SNR $\approx$ 3), and 4U 1850--087 (SNR $\approx$ 2); shown by filled symbols]} lie above the noise curve, and are thus viable candidates for detection. The especially compact sources 4U 1728--34 and 4U 1820--30 ($P_{\text{orb}} \sim 11 \text{ min}$) lie in the most sensitive region for LISA, and could be detected with much shorter observation times $T \ll 4$ yr [see also \cite{chen21} for a discussion on {Taiji and TianQin} sensitivities]. A longer observation time or improved sensitivity would likely be necessary to detect the remaining sources (shown by hollow symbols). However, as noted by \cite{tauris18}, the definition \eqref{eq:chirp} may not be appropriate for systems with active mass transfers, where the companion mass $\Mc$ is effectively dynamical and $\dot{\mathcal{M}} \neq 0$. Had we instead adopted the \emph{dynamical chirp mass} introduced by \cite{tauris18} [see equation (4) therein], we would predict a significantly (factor $\sim 10$ or more) larger effective strain $h_{c}$ for some systems, particularly for XB 1916--053 \citep{iar20} and XTE J1751--305 \citep{rigg11} which have $|\dot{P}_{\text{orb}}| \gtrsim 10^{-11}$. These latter objects may therefore also be detectable by LISA. Overall, the simple analysis conducted in this section demonstrates that UCXBs can be promising candidates for detection by space-borne detectors, as expected \citep{chen20,luo20,huang20}.

\section{UCXBs as LIGO sources}

\begin{table*}
\centering
\caption{Observed and derived properties related to the NSs within the UCXBs considered in Tab. \ref{tab:orbdata}. Spin frequencies, listed in the second column, are deduced from (in order of decreasing robustness): pulsar timing (asterisks), type I X-ray burst tracking (daggers), and QPO frequency differentials (hashes). Surface magnetic field strength maxima $B_{\star}^{\text{max}}$ are estimated from expression \eqref{eq:bmax}. For 4U 0513--40 and 4U 1850--087, where spin frequencies are unavailable, we instead assume $B_{\star}^{\text{max}} = 10^{8.5} \text{ G}$ and estimate $\nu_{\text{spin}}$ from \eqref{eq:bmax}.}
\begin{tabular}{ccccc}
 \hline
 \hline
Source & $\nu_{\text{spin}}$ (Hz) & $B_{\star}^{\text{max}}$ ($\times 10^{8}$ G) & $M_{\star}$ ($M_{\odot}$) & $R_{\star}$ (km) \\
\hline
4U 1820--30$^{\text{a}}$ & $275^{\#}$ & 17.1 & $\sim$ 1.58 & $\sim$ 9.1 \\
IGR J16597--3704$^{\text{b}}$ & $105.2^{*}$ & 14.0 & (1.6) & (10) \\
4U 1728--34$^{\text{c}}$ & $363^{\dag,\#}$ & 3.21 & $\gtrsim$ 1.61 & $\gtrsim$ 9.6 \\
HETE J1900.1--2455$^{\text{d}}$ & $377.3^{*}$ & 2.38 & (1.6) & (10) \\
XB 1916--053$^{\text{e}}$ & $270^{\dag}$ & 5.22 & $\lesssim$ 2.2 & (10) \\
XTE J1807--294$^{\text{f}}$ & $190.6^{*}$ & 16.5 & $\gtrsim$ 1.67 & $\gtrsim$ 8.2 \\
4U 1915--05$^{\text{g}}$ & $366^{\#}$ & 3.14 & (1.6) & (10) \\
4U 0614+091$^{\text{h}}$ & $477^{\dag}$ & 1.74 & $\lesssim 1.6$ & (10) \\
4U 0513--40$^{\text{i}}$ & ($\gtrsim$ 398) & (3.16) & (1.6) & (10) \\
4U 1850--087$^{\text{j}}$ & (403) & (3.16) & $\gtrsim$ 1.23 & $\gtrsim$ 7.16 \\
XTE J1751--305$^{\text{k}}$ & $435.3^{*}$ & 4.48 & (1.6) & (10) \\
XTE J0929--314$^{\text{l}}$ & $185.1^{*}$ & 10.3 & (1.6) & (10) \\
\hline
\hline
\end{tabular}
\\
\footnotesize{$\vphantom{\text{a}}^{\text{a}}$ \cite{white97,guv10} [though cf. \cite{oz16}]. $\vphantom{\text{b}}^{\text{b}}$ \cite{sanna18}. $\vphantom{\text{c}}^{\text{c}}$ \cite{stroh97,shap03}. $\vphantom{\text{d}}^{\text{d}}$ \cite{watts09}. $\vphantom{\text{e}}^{\text{e}}$ \cite{gal01,iar15}. $\vphantom{\text{f}}^{\text{f}}$ \cite{rigg08}; High-mass or large-radius stars preferred \citep{leahy11}. $\vphantom{\text{g}}^{\text{g}}$ \cite{zio99}. $\vphantom{\text{h}}^{\text{h}}$ \cite{klahn06,saz20} [though cf. \cite{stroh08} for a slightly different spin-frequency estimate]. $\vphantom{\text{i}}^{\text{i}}$ Difficult to measure mass or radius \citep{guv12}. Thermonuclear bursts detected at $\approx 1.38$ kHz \citep{bil19}, consistent with rapid rotation. $\vphantom{\text{j}}^{\text{j}}$ \cite{ray04}, assuming a strange star based on resonance absorption features. $\vphantom{\text{k}}^{\text{k}}$ \cite{mark02}; estimates exist for $M_{\star}$ and $R_{\star}$ if the 2002 X-ray outburst is attributable to an $r$-mode, but considered unlikely by \cite{and14}. $\vphantom{\text{l}}^{\text{l}}$ \cite{gal02}.}
\label{tab:nsdata}
\end{table*}

Inferences about the NS primaries within the UCXBs discussed in Sec. 2 have been made from various techniques in the literature. For example, in the case of accreting millisecond-pulsars (AMXPs; 5 of our sources), the NS spin frequencies can be deduced directly from the timing of pulses. In the case of sources exhibiting quasi-periodic oscillations (QPOs; 3 sources) or thermonuclear bursts (3 sources), the spin frequency of the NS can be estimated by assuming that sinusoidal, luminosity fluctuations are due to rotational modulations [see \cite{watts12} for a review]. For systems where measurements of the bursting surface area and redshift are also available, these latter techniques also provide an estimate on the mass-radius relationship of the star in question \citep{van79,oz09,guv10}. Collated in Tab. \ref{tab:nsdata} are parameters relevant to the NSs within the UCXBs listed in Tab. \ref{tab:orbdata}. 

\subsection{Magnetic field estimates}

In general, NSs within active binaries gain angular momentum via accretion torques. Magnetic braking, however, counteracts this spin-up to a degree which depends on the radial extent and thickness of the accretion disc, together with the particulars of the NS magnetic field \citep{ghosh79,white97,bozzo09}. For sufficiently strong fields $B_{\star} \gtrsim 10^{8} \text{ G}$, to-be-accreted material within the volume bounded by the \emph{magnetospheric radius} $R_{\text{m}}$ -- defined as the radius where the ram pressure of infalling matter balances the (outgoing) magnetic pressure -- is funnelled onto the magnetic poles of the NS. In particular, one has $R_{\text{m}} = \xi (B_{\star} R_{\star}^3 )^{4/7} (2 G M_{\star} )^{-1/7} \dot{M}^{-2/7}$, where $\dot{M} \approx L_{X} R_{\star} / G M_{\star}$ denotes the accretion rate and the parameter $0.3 \lesssim \xi \lesssim 1$ accounts for disc-magnetosphere uncertainties, such as the extent of poloidal field-line shearing due to the accretion flow \citep{psal99,and05,pate12}. As the star spins up however, the \emph{co-rotation radius} $R_{\text{co}} = ( G M_{\star} / 4 \pi^2 \nuS^2)^{1/3}$, defined by matching the Keplerian frequency of the disc with $\nuS$, shrinks. If eventually $R_{\text{co}} < R_{\text{m}}$, the rotating magnetosphere will `propeller' plasma back beyond the capture radius \citep{ill75}, effectively halting accretion. Observations of recurrent thermonuclear activity therefore require $R_{\text{co}} \geq R_{\text{m}}$, with upper-limit strength $B^{\text{max}}_{\star}$ obtained at equality, viz.
\begin{equation} \label{eq:bmax}
\begin{aligned}
B_{\star}^{\text{max}} \approx& \, 2.8 \times 10^{8} \left( \frac {500 \text{ Hz}} {\nuS} \right)^{7/6} \left( \frac {M_{\star}} {1.6 M_{\odot}} \right)^{1/3} \\
&\times \left( \frac{L_{X}} {10^{37} \text{ erg s}^{-1}} \right)^{1/2} \left( \frac {10^{6} \text{ cm}} {R_{\star}} \right)^{5/2} \text{ G},
\end{aligned}
\end{equation}
for a canonical choice $\xi = 0.8$. 

Inferred values of $B_{\star}^{\text{max}}$ from equation \eqref{eq:bmax} are listed in the third column of Tab. \ref{tab:nsdata}. For the two systems 4U 0513--40 and 4U 1850--087 (which reside in globular clusters, thereby making it difficult to detect outbursts due to a wealth of confusion noise) where spin frequencies are unavailable, we instead assume that $B_{\star}^{\text{max}} = 10^{8.5} \text{ G}$ and infer the spin frequency from expression \eqref{eq:bmax}; we note that \cite{bil19} reports a detection of $\gtrsim$ kHz QPOs from 4U 0513--40, supporting the possibility of a rapidly rotating NS.

\subsection{Continuous gravitational waves}

The observation that NSs within binaries tend to have saturated spin frequencies $\nuS \lesssim 650 \text{ Hz}$, regardless of system age, leads to the suggestion that additional means of spin-down beyond magnetic braking are operational in some systems \citep{bild98,chak03}. In particular, the observed spin frequencies reported in Tab. \ref{tab:nsdata} could be reconciled with equilibrium values $\gg 500 \text{ Hz}$ if the stars housed a sufficiently large, time-varying quadrupole moment \citep{and99}, at least transiently \citep{hmount15}, so as to augment the spin-down rate. Theoretical considerations along these lines have motivated searches for GWs from binary systems [e.g., \cite{suva16,middle20}], including targeted searches for XTE J1751--305 \citep{mead17}. Furthermore, much like LMXBs \citep{shibaz89}, the surface magnetic field strengths of UCXBs are generally quite low $(B_{\star} \lesssim 10^{9} \text{ G})$, as can be seen from expression \eqref{eq:bmax} and Tab. \ref{tab:nsdata}, relative to expectations from population synthesis models. This suggests that the stars may have undergone magnetic burial during their binary lifetimes \citep{payne04,suvm20} (see Sec. 3.3), and house accretion-built magnetic mountains which produce the necessary quadrupole moments \citep{melp05,pri11}. Other promising means of generating large quadrupole moments, potentially detectable by aLIGO or the next-generation ET, come from toroidal fields in superconducting cores (Sec. 3.4) and $r$-mode oscillations (Sec. 3.5).

Based on the three mechanisms for generating quadrupole moments discussed above, we can estimate the extent to which the UCXB sources considered in Tab. \ref{tab:orbdata} could be detectable as continuous GW sources. For a fully coherent search over time $T$, a ground-based interferometer can detect a signal of amplitude $h_{0} \approx 11.4 \sqrt{ S_{n} / T}$ with $90\%$ confidence \citep{watts08}. In general, a NS with mass quadrupole moment $Q_{22}$ emits gravitational waves at frequency $f_{\text{GW}} = 2 \nuS$ and amplitude\footnote{{We caution the reader that the `angle-averaged' amplitude ($h_{a}$) used by \cite{ush00} differs from the intrinsic amplitude ($h_{0}$) more commonly used today through $h_{0} = \sqrt{5/2} h_{a}$; see \cite{aasi14}.}} [adopting the conventions of \cite{ush00}]
\begin{equation} \label{eq:massquadamp}
\hspace{-0.064cm} h_{0} \approx 1.4 \times 10^{-26} \left( \frac {Q_{22}} {10^{38} \text{ g cm}^2} \right) \left( \frac{\nuS} {500 \text{ Hz}} \right)^{2} \left( \frac {10 \text{ kpc}} {d} \right).
\end{equation}
The value of quadrupole moment $Q_{22}$ is of course the primary uncertainty in expression \eqref{eq:massquadamp} in most cases. For mass or current quadrupoles induced by mode oscillations, the system instead radiates at the (inertial-frame) mode frequency $\omega_{i}$ with an amplitude that depends on the mode eigenfunction \citep{and99,and01} (see Sec. 3.5 for details in the $r$-mode case).

{While realistic models of accretion flow are complicated, the spin-up torque exerted on the star can be generally approximated as $N_{a} \approx \dot{M} \left( G M_{\star} R_{\text{m}} \right)^{1/2}$ \citep{ghosh79,bozzo09,chens20}. An observational upper limit for the quadrupole moment(s) can then be set by assuming that GW (as opposed to electromagnetic) braking dominates the rotational evolution \citep{ush00}. Setting $2 \pi I_{0} \dot{\nu} = N_{a} - N_{\text{GW}}$, where $N_{\text{GW}} = 2^{13} G \pi^6 Q_{22}^2 \nu_{\text{spin}}^{5} / 75 c^5$ for mass quadrupoles (see also Sec. 3.5), the NS spin frequency derivative reads}
\begin{equation} \label{eq:sdmax}
\begin{aligned}
\dot{\nu} \approx&\, 1.4 \times 10^{-13} \left( \frac {10^{45} \text{ g cm}^2} {I_{0}} \right) \Bigg[ \left( \frac{L_{X}} {10^{37} \text{ erg s}^{-1}} \right)^{6/7} \\
& \times \left( \frac{ B_{\star}} {10^{8} \text{ G}} \right)^{2/7} \left( \frac {R_{\star}} {10^{6} \text{ cm}} \right)^{12/7} \left( \frac {1.6 M_{\odot}} {M_{\star}} \right)^{3/7} \\
 &- 106 \left( \frac {Q^{\text{max}}_{22}} {10^{38} \text{ g cm}^2} \right)^2 \left( \frac{\nuS} {500 \text{ Hz}} \right)^{5} \Bigg] \text{ Hz s}^{-1},
\end{aligned}
\end{equation}
{for moment of inertia $I_{0} \approx 2 M_{\star} R_{\star}^2/5 \gtrsim 10^{45} \text{ g cm}^2$ and magnetic field strength $B_{\star} \sim 10^{8} \text{ G}$ (greater strengths imply weaker maximum quadrupoles).} For systems in spin equilibrium, where the terms inside the square brackets cancel out, one finds $10^{37} \lesssim Q^{\text{max}}_{22}/( \text{g cm}^{2}) \lesssim 10^{38}$ for canonical parameters \citep{bild98}. Furthermore, there is a maximum quadrupole moment $Q^{\text{elastic}}_{22}$ that an elastic NS crust can support. For a cold and catalysed crust, this maximum is thought to lie in the range $10^{39} \lesssim Q^{\text{elastic}}_{22} / (\text{g cm}^{2}) \lesssim 10^{40}$ \citep{mcd13}, though scales with the stellar compactness. For an accreted crust, $Q^{\text{elastic}}_{22}$ is reduced by a factor $\sim 2$ \citep{hask06}. In any case, $Q^{\text{elastic}}_{22} \gg Q_{22}$ for all systems considered here.

\begin{figure}
\hspace{-0.3cm}\includegraphics[width=0.5\textwidth]{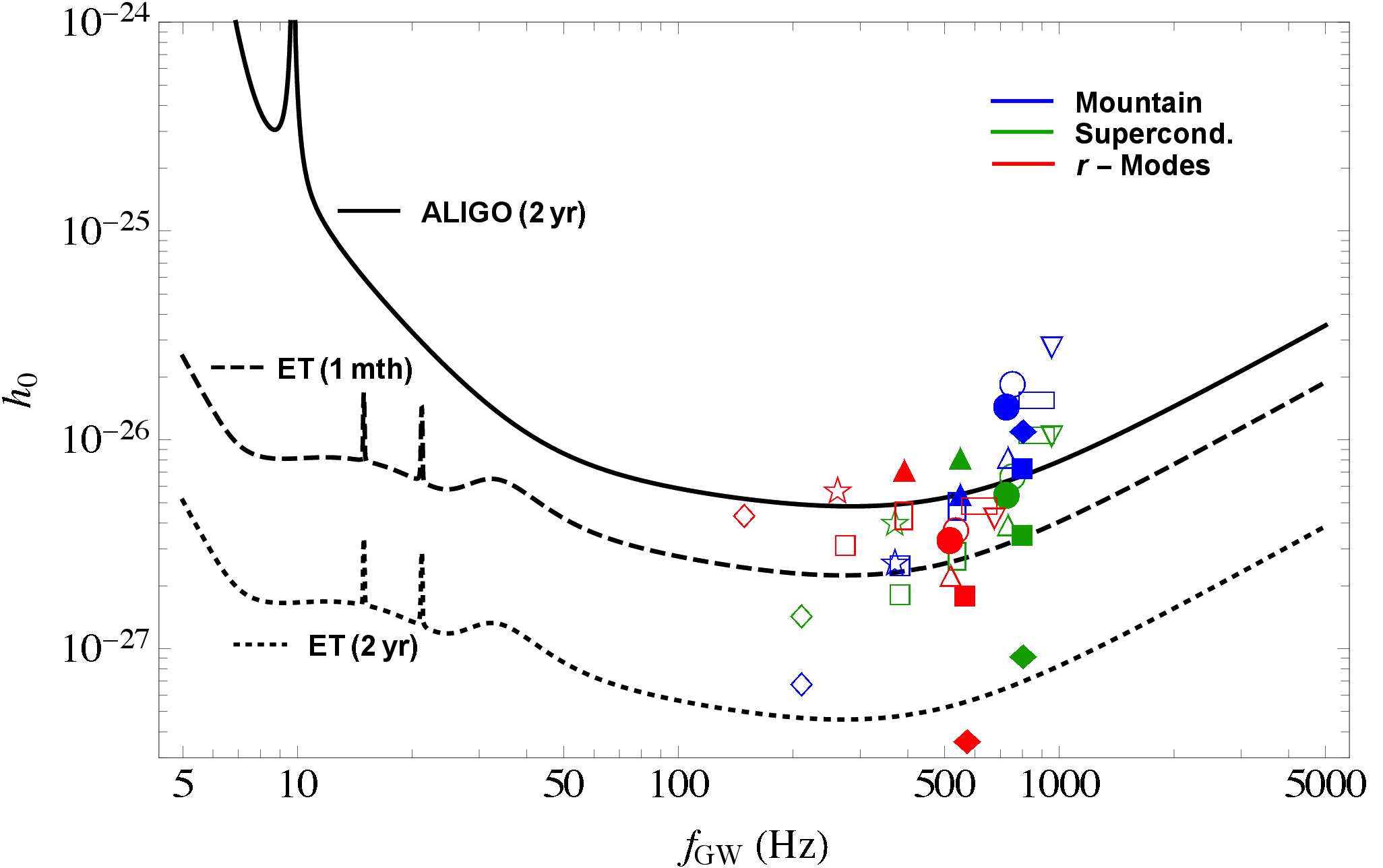}
\caption{Intrinsic GW strain $h_{0}$ as a function of $\fGW$ for the twelve UCXB sources listed in Tab. \ref{tab:orbdata}, estimated assuming quadrupole moments induced by magnetic mountains (blue; Sec. 3.3), toroidal fields in superconducting stars (green; Sec. 3.4), and $r$-mode oscillations (red; Sec. 3.5). The particular symbols correspond to the same sources in the plot legend of Fig. \ref{fig:lisastrains}; e.g., hollow circles correspond to HETE J1900.1--2455. Overplotted are the sensitivity curves for $\text{aLIGO}^{3}$ (solid, $T = 2$ yr) and $\text{ET-D}^{4}$ (dashed, $T = 1$ month; dotted, $T = 2$ yr) assuming a phase-coherent search of duration $T$ \citep{watts08}.} \label{fig:ligostrains}
\end{figure}

\begin{figure}
\hspace{-0.2cm}\includegraphics[width=0.5\textwidth]{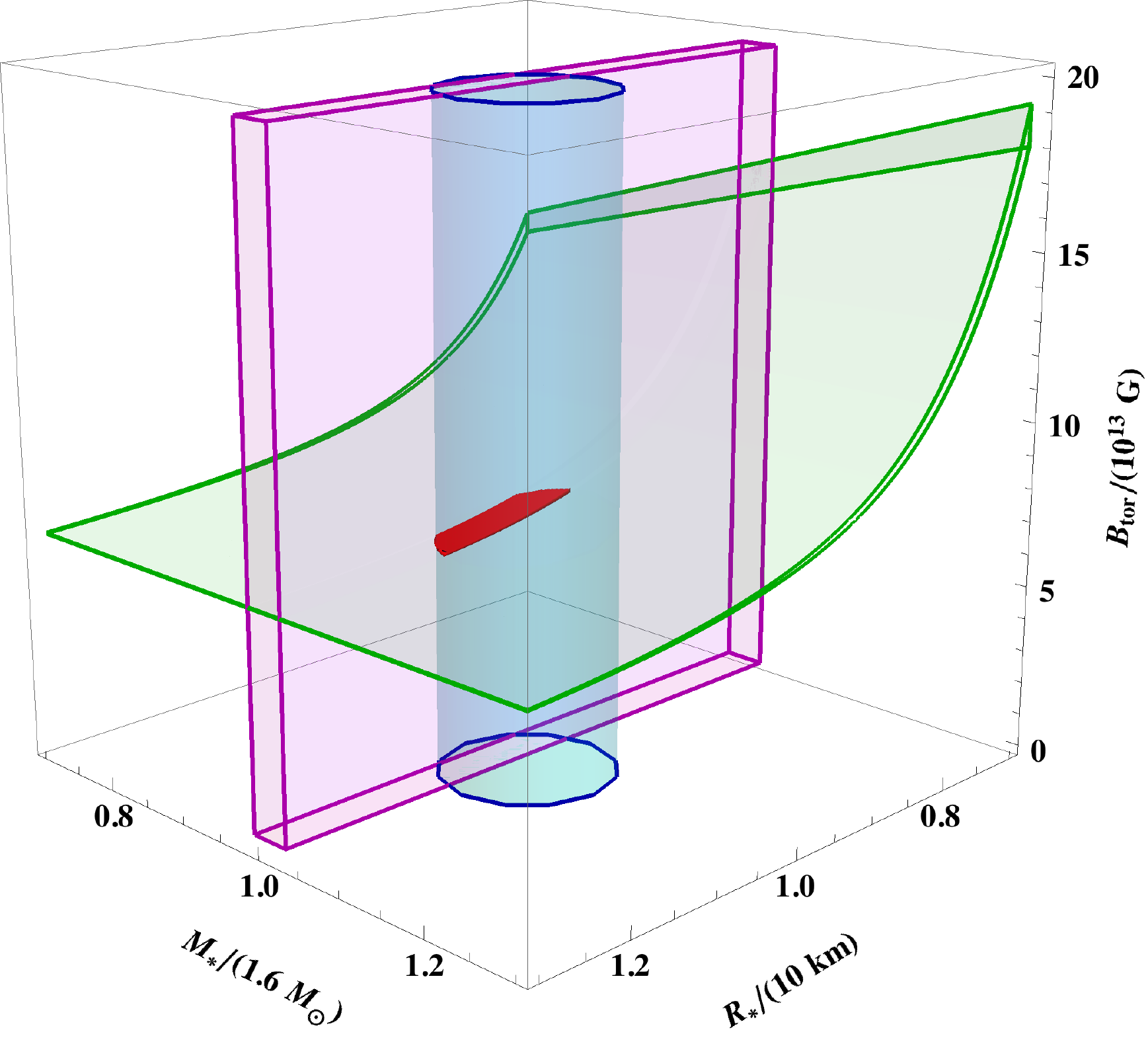}
\caption{A depiction of hypothetical constraints on the $M_{\star}$-$R_{\star}$-$B_{\text{int}}$ configuration space of a NS following combined dual-line and thermonuclear burst detections. (i) Magenta cuboid: orbital GWs observed by LISA, Taiji, or TianQin could be expected to place tight constraints on $M_{\star}$ for high SNRs \protect\citep{chen21}, especially in systems with WD companions. In this case, no constraints are placed on $B_{\text{int}}$ or $R_{\star}$, so a thin cuboid centered around the true mass ($M_{\star} = 1.6 M_{\odot}$, say) is allowed. (ii) Green sheet: continuous GWs observed by aLIGO or ET would bound the quadrupole moment $Q_{22}$ within some tolerance. For the superconductor expression \eqref{eq:superquad}, this implies a constraint on $B_{\text{tor}} R_{\star}^6/ M_{\star}$. In this case, because of the mutual dependence on all three parameters, the allowed region resembles a curved sheet with thickness that is proportional to measurement uncertainties. (iii) Blue cylinder: for thermonuclear flashes where the apparent emitting area and peak flux achieved during photospheric bursts can be measured, the mass-radius relationship of the star can be constrained \protect\citep{guv10}, as depicted by the blue circle in the $M_{\star}$-$R_{\star}$ subplane. In this case, no constraint on the $\boldsymbol{B}$ field is obtained, so the circle stretches into a cylinder along the $B_{\text{tor}}$ axis. The intersection between surfaces (i)--(iii) is shown in red.} \label{fig:confspace}
\end{figure}

Figure \ref{fig:ligostrains} shows {theoretical} \emph{upper limits} for the GW strains associated with magnetic mountains (Sec. 3.3; blue symbols), toroidal fields in a superconducting star (Sec. 3.4; green symbols), and $r$-modes (Sec. 3.5; red symbols). In particular, the hollow or filled symbols correspond to the same sources as depicted in Fig. \ref{fig:lisastrains}; the source 4U 0513--40 is denoted by filled squares, for example. Overplotted in Fig. \ref{fig:ligostrains} are the aLIGO\footnote{aLIGO data from \url{https://dcc.ligo.org/public/0149/T1800044/005}} and ET-D\footnote{ET data from \url{http://www.et-gw.eu/index.php/etsensitivities}} sensitivity curves, assuming a generous observation time of $T = 2$ yr or a shorter run of $T = 1$ month. Recalling that the filled symbols correspond to those sources which are likely detectable by LISA within 4 years of observation time, we see that there are at least two candidates for dual-line detection by aLIGO: 4U 1728--34 (solid circles) and 4U 1820--30 (solid upper-triangles), with 4U 1850--087 (solid diamonds) being a marginal third candidate. These former sources are rapidly rotating ($\nuS = 363$ Hz and $\nuS = 275$ Hz, respectively), have relatively high X-ray luminosities ($L_{X} \sim 5 \times 10^{36} \text{ erg s}^{-1}$ and $L_{X} \sim 6 \times 10^{37} \text{ erg s}^{-1}$, respectively), and are active bursters \citep{saz20}. A detection of orbital \emph{and} continuous GWs from these systems, which already have relatively tight mass and radius constraints from X-ray burst tracking \citep{stroh97,shap03,gal10,guv10}, may provide unparalleled information concerning the EOS of nuclear matter (since $h_{c}$ and $h_{0}$ are tied to $M_{\star}$ and $R_{\star}$), magnetic substructure (since $h_{0}$ is tied to $B_{\text{int}}$), and accretion dynamics (since $h_{0}$ is tied to $\dot{M}$) \citep{hmount15,glamp18}. {An illustration of how a combined observation from these three channels can be used to pinpoint NS properties is drawn in Figure \ref{fig:confspace}.}

Overall, however, emphasising that we have considered upper limits for the quadrupole moments (see Secs. 3.3--3.5), the outlook for detection of continuous GWs from UCXB sources with aLIGO is relatively pessimistic (with the possible exception of the two sources detailed above, HETE J1900.1--2455, and 4U 0614+091), {especially if the system is in spin equilibrium}. For XTE J1751--305 (hollow horizontal-rectangles), \cite{mead17} placed upper limits on the continuous strain of $h_{c} \leq 4.7 \times 10^{-24}$ at $\fGW \gtrsim 4 \nuS/3$ and $h_{c} \leq 7.8 \times 10^{-24}$ at $\fGW = 2 \nuS$. These values are roughly two orders of magnitude higher than those predicted here, though we note that \cite{mead17} made use of data from LIGO's science run 6 (less sensitive than design sensitivity by roughly an order of magnitude), and did not have the luxury of observing for 2 yr. However, pulse timing reveals a secular spindown of $\dot{\nu}^{\text{J1751}}_{\text{spin}} \sim -5.5 \times 10^{-15} \text{ Hz s}^{-1}$ \citep{rigg11}, which, {from expression \eqref{eq:sdmax}}, limits the maximum mass quadrupole moment to $Q^{\text{max}}_{22} \approx 3.1 \times 10^{36} \text{ g cm}^{2}$; greater quadrupole moments imply faster spin down \citep{bild98}. We likely overestimate $h_{0}$ for this object therefore, and ET would be required. By contrast, during an active phase for XTE J0929--314 (hollow stars), \cite{gal02} found $\dot{\nu}^{\text{J0929}}_{\text{spin}} \sim -9.2 \times 10^{-14} \text{ Hz s}^{-1}$, which sets an upper limit that exceeds our estimates, $Q^{\text{max}}_{22} \approx 1.7 \times 10^{38} \text{ g cm}^{2}$ [see also \cite{middle20}]. {More recently, the Einstein@Home project was able to exclude the existence of nearby $(d \lesssim 100 \text{ pc})$ NSs with quadrupole moments $\gtrsim 10^{38} \text{ g cm}^2$ for rotation rates $10 \leq \nu_{\text{spin}} / \text{Hz} \leq 293$ \citep{stelt21}.} In any case, our simple analysis indicates that virtually all sources may be detectable by ET, even for observation times $T \ll 2$ yr in some cases (see dashed curve in Fig. \ref{fig:ligostrains}), which further strengthens the case for the design and construction of next-generation detectors.

\subsection{Magnetic mountains}

NSs within actively accreting binaries are expected to undergo episodes of magnetic burial, where infalling matter is guided onto the polar cap(s) by the NS magnetic field, eventually forming mountain-like accretion columns which are supported by the Lorentz force from the compressed magnetic field \citep{brown98,payne04,pri11}. The compression of magnetospheric plasma leads to a reduction in the global dipole moment of the NS [potentially explaining low $B$-field observations \citep{shibaz89}], while simultaneously generating a mass quadrupole moment that leads to the emission of GWs \citep{melp05}. UCXBs that exhibit thermonuclear activity with recurrence times that are too short for many theoretical ignition models ($\lesssim 10$ min) are especially promising candidates for housing accretion-built mountains atop their surface, since buried, multipolar fields may assist in the magnetic confinement of local patches of fuel \citep{suvm19,suvm20,gall21}.

From an empirical standpoint, \cite{shibaz89} proposed the relationship
\begin{equation} \label{eq:shibaz}
\mu = \mu_{i} \left(1 + M_{a} / M_{c} \right)^{-1}
\end{equation}
for LMXB dipole moment $\mu$ (with pre-accretion value $\mu_{i}$), accreted mass $M_{a}$, and `critical' mass $M_{c}$. For low values of $M_{a}$ relative to $M_{c}$, numerical simulations of Grad-Shafranov equilibria validate the relationship \eqref{eq:shibaz} and predict a mass quadrupole moment $Q_{22} \sim 10^{45} A \left(M_{a} / M_{\odot} \right) \left( 1+ M_{a}/M_{c} \right)^{-1} \text{ g cm}^{2}$, where $A$ is an order-unity factor that depends on the crustal EOS and accretion geometry \citep{melp05}. On the other hand, for $M_{a} \gtrsim M_{c}$, the simple estimate \eqref{eq:shibaz} breaks down: ideal magnetohydrodynamic (MHD) simulations suggest instead that burial tends to reach levels of $10^{-2} \lesssim \mu/ \mu_{i} \lesssim 10^{-1}$ (again depending on the pre-accretion magnetic field strength and crustal EOS) for $M_{a} \lesssim 10 M_{c}$ \citep{payne04,pri11,hmount15,suvm19}, though long-term Ohmic decay may reduce the global dipole moment by up to factors of $\sim 10^{4}$ \citep{vig09}. Burials of order $\gtrsim 10^{2}$ are consistent with cyclotron-line measurements made in X-ray binaries, which show that accreting NSs can possess local fields that are a couple orders of magnitude greater than those inferred from global measurements \citep{stau19}.

Simulations of magnetic equilibria for $M_{a} \gtrsim M_{c}$ suggest that the mass quadrupole moment saturates at 
\begin{equation} \label{eq:mountquad}
Q_{22} = A(M_{\star},R_{\star},\mu_{i},\dot{M},M_{a}) \times 10^{38} \text{ g cm}^{2},
\end{equation}
for geometric factor $0.1 \lesssim A \lesssim 10$ that depends on the EOS and accretion physics [see \cite{pri11} and \cite{hmount15} for details]. The estimate \eqref{eq:mountquad} could, however, potentially be reduced by a factor $\gtrsim 2$ by time-dependent Parker \citep{vig09} or ballooning \citep{mukh13a,mukh13b} instabilities. Using ideal-MHD simulations which allowed for accreted material to sink into the lower density substrate, \cite{wette10} found that for $M_{a} \sim 0.12 M_{\odot} \gg M_{c}$, $Q_{22}$ is increased by a factor $\sim 2$ relative to the $M_{a} \gtrsim M_{c}$ value when considering an isothermal crust, which is a fair description for low accretion rates $\dot{M} \lesssim 10^{-10} M_{\odot} \text{ yr}^{-1}$ ($L_{X} \lesssim 10^{36} \text{ erg s}^{-1}$) \citep{fuji}. We stress though that cases with $M_{a} \gg M_{c}$ (the regime most likely applicable to old UCXB systems) are numerically inaccessible at present for realistic, crustal EOS, and therefore the above estimates should be treated as upper limits. 

For the mountain GW strains (blue symbols) presented in Fig. \ref{fig:ligostrains}, we therefore take a canonical mass quadrupole moment $Q_{22} \approx 10^{38} \text{ g cm}^{2}$ [i.e., $A=1$ in expression \eqref{eq:mountquad}], and $h_{0}$ is computed from \eqref{eq:massquadamp}. 

\subsection{Superconducting cores with toroidal fields}

Mature NSs are generally expected to be in both a superfluid and superconducting state \citep{baym69}. In particular, when the core temperature falls below a critical threshold [$T_{\text{crit}} \sim 5 \times 10^{9}$ K; \citep{gas11}] due to neutrino cooling, neutrons and protons become superfluid and superconducting, respectively, as the formation of Cooper pairs becomes energetically favourable. This theoretical consideration is generally supported by the observation of pulsar glitches \citep{hm15} and NS cooling curves \citep{and21}. In any case, if the protons in the interior form a type II superconductor, then the magnetic field is quantized into flux tubes and the magnetic force is a tension force associated with the (pinned) flux tubes, with critical strengths $10^{15} \lesssim H_{c1} \lesssim 10^{16}$ G \citep{gas11}, rather than the Lorentz force \citep{rud91,glamp18}. This tension force therefore scales with the effective energy $\sim B H_{c1} / ( 8 \pi)$, which can be large even in stars with relatively low magnetic field strengths.

Although keeping in mind the uncertainties detailed in the previous section, consider a scenario where the internal magnetic field remains at some pre-accretion strength over the UCXB lifetime, while the global dipole is reduced through burial or Ohmic decay by a factor $b \sim 10^2$. The internal, \emph{poloidal} magnetic field will thus be $\sim 10^{2}$ times larger than the external dipole field inferred by expression \eqref{eq:bmax}. However, it is well known that purely poloidal magnetic fields are unstable \citep{tay73}, and $\boldsymbol{B}$ must contain a (possibly large) toroidal component. Although only applicable to single-fluid stars with normal magnetic fields, the stability analyses of \cite{brai09} and \cite{akgun13} suggest that barotropic and non-barotropic NSs can support toroidal fields that are $\lesssim 10^{3}$ times stronger than their respective poloidal components [though cf. \cite{lander13}]. Differential rotations induced by dynamo activity \citep{td93} or $r$-mode oscillations (see Sec. 3.5) shortly after birth or later in life \citep{rls00,rez01} may work towards the winding-up of poloidal field lines, which could potentially generate toroidal fields of this strength. Putting this together, we have that, for the canonical estimate in \eqref{eq:bmax}, NSs in UCXBs may contain hidden toroidal fields of upper-limit strength $B_{\text{tor,max}} \lesssim 10^{14} \text{ G}$. 
	
For a toroidal field of volume-averaged strength $\langle B_{\text{tor}} \rangle = \lambda b B_{\star}^{\text{max}}$, where $\lambda \gg 1$ is a parameter that quantifies the strength of the toroidal field relative to the poloidal one, the mass quadrupole moment associated with a superconducting star can be estimated as \citep{cutler02}
\begin{equation} \label{eq:superquad}
\begin{aligned}
Q_{22} \approx&\, 1.2 \times 10^{38} \left( \frac{\lambda b B_{\star}^{\text{max}}} {10^{14} \text{ G}} \right) \left( \frac {H_{c1}} {10^{15} \text{ G}} \right) \left( \frac {1.6 M_{\odot}} {\Ms} \right)^{2} \\
&\times \left( \frac {I_{0}} {10^{45} \text{ g cm}^2} \right) \left( \frac {\Rs} {10^{6} \text{ cm}} \right)^{4} \text{ g cm}^2 .
\end{aligned}
\end{equation}
Note that the same quadrupole moment \eqref{eq:superquad} would be predicted for a toroidal field with a more modest strength $B_{\text{tor}} \sim 10^{13} \text{ G}$ (i.e., $\lambda \sim 10^{2}$) if instead we took $H_{c1} \sim 10^{16} \text{ G}$, a value which is allowed within the theoretical uncertainties \citep{gas11}. The same applies for $B_{\star} \sim 0.1 B_{\star}^{\text{max}}$ and $H_{c1} \sim 10^{16} \text{ G}$. From expression \eqref{eq:massquadamp}, we can calculate the GW strain associated with the superconductor quadrupole moment \eqref{eq:superquad}; the results are shown by the green symbols in Fig. \ref{fig:ligostrains}.

\subsection{$r$-mode oscillations}

Another viable candidate for generating large GW strains in millisecond NSs comes from $r$-mode pulsations \citep{and99,hask15}. Leading-order $(\ell = m = 2)$ inertial modes are known to be retrograde in the rotating frame of the star but prograde in the laboratory frame. They are therefore subject to the secular Chandrasekhar-Friedman-Schutz (CFS) instability, in the sense that gravitational-radiation reaction tends to amplify the modes rather than damp them out \citep{chand70,fs78}. Viscosity acts to suppress the modes however, and so a competition ensues between the GW and viscous timescales, defining the so-called $r$-mode `instability window' that depends on the spin and temperature of the NS \citep{ands99,lev99,and01}. As noted in Sec. 3.4, the NSs within UCXBs are likely to be relatively cool [$T_{\text{core}} \lesssim 10^{8} \text{ K}$; \cite{mahm13}], which may put them within the instability window for spins $\gtrsim 300 \text{ Hz}$ [see Figure 4 of \cite{hask15}]; however, the exact geometry of the window is sensitive to the extent of viscous damping taking place at the base of the crust, which is highly uncertain \citep{andj00,kokk16,glamp18}.

For $r$-mode pulsations, GWs are emitted at the inertial-frame frequency of the modes \citep{owen10,and14}. For the $\ell = m =2$ modes, one has 
\begin{equation} \label{eq:rmodefreq}
\fGW = \omega_{i} \approx \left( \frac {4} {3} + \frac {16 G \Ms} {45 \Rs c^2} \right) \nu_{\text{spin}},
\end{equation}
where the second term within the parentheses, derived by \cite{lock03}, is a post-Newtonian correction that accounts for the compactness of the star. GW generation in this case is driven by the current-quadrupole moment, and the intrinsic amplitude reads [see, e.g., equation (31) in \cite{glamp18}]
\begin{equation} \label{eq:rmodeamp}
\begin{aligned}
h_{0} \approx&\, 5.7 \times 10^{-28} \left( \frac {\alpha_{r}} {10^{-6}} \right) \left( \frac{ \omega_{i}} {500 \text{ Hz}} \right)^{3} \\
&\times \left( \frac {10 \text{ kpc}} {d} \right) \left( \frac {\Ms} {1.6 M_{\odot}} \right) \left( \frac{\Rs} {10^{6} \text{ cm}} \right)^{3},
\end{aligned}
\end{equation}
where $\alpha_{r}$ denotes the $r$-mode amplitude, the limiting physics of which is not fully understood [see, e.g., \cite{hask14}]. A nominal bound can be set by assuming accretion-torque balance \citep{brown00} {[cf. equation \eqref{eq:sdmax}]}, viz.
\begin{equation} \label{eq:satamp}
\alpha_{r} \approx 1.4 \times 10^{-6} \left( \frac {L_{X}} {10^{37} \text{ erg s}^{-1}} \right)^{1/2} \left( \frac {500 \text{ Hz}} {\nuS} \right)^{7/2}.
\end{equation}
However, as discussed by \cite{owen10}, the above estimate may be pessimistic for some sources since nonlinear couplings and sporadic accretion episodes can allow for large, transient current quadrupoles. On the other hand, in addition to deep crustal heating induced by accretion itself \citep{haen03}, viscous shearing from $r$-mode oscillations may set a thermal limit $\alpha_{\text{th}}$: the sum of the thermal photon and neutrino luminosities cannot be less than that induced by $r$-mode heating if the system is to avoid thermogravitational runaway \citep{lev99,andj00}. \cite{mahm13} [see also \cite{sch17} and \cite{boz20}] found that $0.04 \lesssim \alpha_{\text{th}}/ \alpha_{r} \lesssim 0.8$ for LMXBs where quiescent luminosity data is available, implying that \eqref{eq:satamp} may be an over-estimate. Furthermore, amplitudes \eqref{eq:satamp} assume no net spin-down but also insignificant magnetic braking -- likely unrealistic for most sources.

Characteristic wave strains associated with $r$-modes, evaluated from expression \eqref{eq:rmodeamp} with amplitudes \eqref{eq:satamp}, are shown by the red symbols in Fig. \ref{fig:ligostrains}. These cases are generally more pessimistic than their mass-quadrupole counterparts, though the detectability of 4U 1820--30 (solid upper-triangles) via $r$-modes is competitive with that predicted by the superconducting model. XTE J0929--314 (hollow stars) appears to be a notable exception, as our estimates suggest this source may be detectable by aLIGO, however this object likely spins too slowly to reside within the instability window \citep{hask15}; see also \cite{and14} for a discussion regarding XTE J1751--305 (horizontal rectangles).

\section{Discussion}

Ultra-compact binaries with sub-hour orbital periods are expected to emit GW radiation in the $\sim$ mHz range, which lies in the peak sensitivity trough of the space-borne GW detectors LISA, {Taiji, and TianQin} \citep{rob19,luo20,huang20}. Using evolutionary models, \cite{tauris18} found that the masses of companion stars in detached post-LMXBs, that are theoretically detectable in orbital GWs, tend to lie in an extremely narrow range around a function of $P_{\text{orb}}$. This suggests that the only major uncertainties within the chirp mass \eqref{eq:chirp} pertain to the primary, so that a detection of orbital GWs from these systems would provide strong constraints on the mass of the NS: \cite{tauris18} goes on to suggest that LISA could place constraints on $M_{\star}$ at the $\sim 4\%$ level. Further evolutionary modelling by \cite{chen20} suggests that up to $\approx 320$ LISA-resolvable NS-WD binaries, {evolving from the NS--main-sequence channel, could reside in the Galaxy [see also \cite{hein13,seng17}]}. In this work, we identified 4 known UCXB systems (4U 1820--30, 4U 1728--34, 4U 0513--40, and 4U 1850--087) that may be detectable within at most 4 years of observation time (see Fig. \ref{fig:lisastrains}).

In addition to being sources of \emph{orbital} GWs at $\sim$ mHz frequencies, UCXB systems are also promising sources of \emph{continuous} GWs at $\lesssim$ kHz frequencies \citep{tauris18}. In particular, a proposed solution to the observational puzzle of capped spin frequencies in active binaries is that gravitational radiation-reaction offsets the accretion-driven growth of $\nuS$ \citep{bild98,and01,chak03}. For example, large mass quadrupole moments can be generated by accretion-built magnetic mountains \citep{melp05,suvm19} or internal toroidal fields in the cores of mature and cool NSs \citep{cutler02,gas11}. Non-negligible current quadrupoles can similarly be generated by $r$-mode oscillations \citep{and99,and01}. Considering these three mechanisms, we estimated upper-limits to the continuous GW strain from twelve UCXB systems (Fig. \ref{fig:ligostrains}), and found that at least two sources are potentially detectable by both LISA \emph{and} aLIGO simultaneously: 4U 1728--34 and 4U 1820--30. Dual-line detections from these sources would give two data points in their respective $M_{\star}$-$R_{\star}$-$B_{\text{int}}$ spaces, from which stringent conclusions concerning the nuclear EOS and internal magnetic field structure can be drawn {(green and magenta surfaces in Fig. \ref{fig:confspace})}. Such studies may be naturally supplemented by electromagnetic observations: many UCXB systems are X-ray bright and active (emitting thermonuclear bursts and/or QPOs), so that masses and radii [e.g., 4U 1728--34 \citep{shap03} and 4U 1820--30 \citep{guv10}] can be independently constrained {(blue surface in Fig. \ref{fig:confspace}). It is hoped therefore that this work will encourage future GW searches to be targeted at UCXBs.}

Finally, it is worth pointing out that there are many UCXB systems that could serve as LISA sources that are not included here (see Tab. \ref{tab:orbdata}). One interesting example is 4U 1626--67, which has an orbital period of $\sim$ 42 minutes (possibly visible by LISA), though is a very slowly rotating system with $\nuS \approx 0.13$ Hz \citep{brown98} (invisible to aLIGO). Cyclotron-line measurements for this object imply that the NS has a relatively strong magnetic field, $B \gtrsim 3 \times 10^{12}$ G \citep{orl98}. If cyclotron features were observed in an UCXB, in addition to dual-line and thermonuclear burst measurements, one could constrain not only $M_{\star}$, $R_{\star}$, and $B_{\text{int}}$, but the magnetic field \emph{geometry} also (e.g., poloidal-to-toroidal strength ratio $\lambda$). Other potential dual-line sources are the nearby ($d \lesssim$ 5 kpc and $d \sim 3.3$ kpc, respectively) X-ray bursters 2S 0918--549 and 4U 1543--624 --- UCXBs with $\approx 17.4$ and $\approx 18.2$ minute orbital periods, respectively \citep{zhong11,lud19} --- though little is known about the NSs. \cite{mason20} recently reported the discovery of X-ray pulsations in IGR J17494--3030, inferring that the source is a $\lesssim 3$ ms AMXP in a $\lesssim 75$ min binary. A secular spindown of this object was measured as $\dot{\nu}^{\text{J17494}}_{\text{spin}} = -2.1 \times 10^{-14} \text{ Hz s}^{-1}$, implying a maximum mass quadrupole $Q^{\text{max}}_{22} \approx 8.9 \times 10^{36} \text{ g cm}^{2}$ \citep{bild98}, potentially detectable by ET. The donor star is quite light however, $M_{\text{comp}} \sim 10^{-2} M_{\odot}$ \citep{mason20}, and the source is likely out of view for LISA.

\section*{Acknowledgements}
I would like to thank Kostas Kokkotas and Kostas Glampedakis for very helpful discussions. {I also thank the anonymous referee for providing helpful feedback, which considerably improved the quality of the manuscript.} This work was supported by the Alexander von Humboldt Foundation. The author is a member of the LISA consortium.

\section*{Data availability}
Observational data used in this paper are quoted from the cited works. Data generated from computations are reported in the body of the paper. Additional data can be made available upon reasonable request.


\end{document}